# Convection-Enhanced Biopatterning with Recirculation of Hydrodynamically Confined Nanoliter Volumes of Reagents

Julien Autebert,[†] Julien F. Cors,[†] David P. Taylor, and Govind V. Kaigala[*]

IBM Research—Zurich, Säumerstrasse 4, 8803 Rüschlikon, Switzerland

*Supporting Information*

**ABSTRACT:** We present a new methodology for efficient and high-quality patterning of biological reagents for surface-based biological assays. The method relies on hydrodynamically confined nanoliter volumes of reagents to interact with the substrate at the micrometer-length scale. We study the interplay between diffusion, advection, and surface chemistry and present the design of a noncontact scanning microfluidic device to efficiently present reagents on surfaces. By leveraging convective flows, recirculation, and mixing of a processing liquid, this device overcomes limitations of existing biopatterning approaches, such as passive diffusion of analytes, uncontrolled wetting, and drying artifacts. We demonstrate the deposition of analytes, showing a 2- to 5-fold increase in deposition rate together with a 10-fold reduction in analyte consumption while ensuring less than 6% variation in pattern homogeneity on a standard biological substrate. In addition, we demonstrate the recirculation of a processing liquid using a microfluidic probe (MFP) in the context of a surface assay for (i) probing 12 independent areas with a single microliter of processing liquid and (ii) processing a 2 mm$^2$ surface to create 170 antibody spots of 50 × 100 $\mu$m$^2$ area using 1.6 $\mu$L of liquid. We observe high pattern quality, conservative usage of reagents, micrometer precision of localization and convection-enhanced fast deposition. Such a device and method may facilitate quantitative biological assays and spur the development of the next generation of protein microarrays.

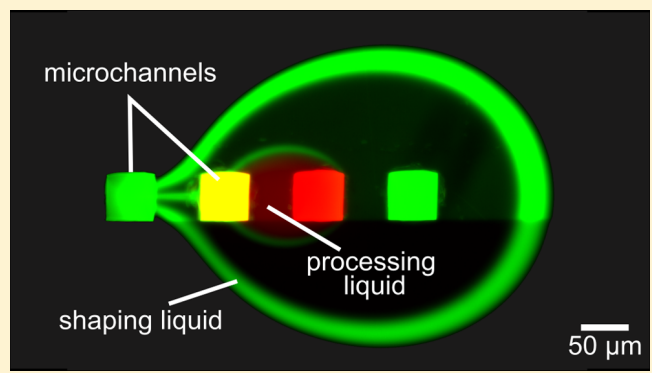

Patterning and immobilization of chemicals, proteins, or biomolecules on surfaces are central to surface biological assays[1−5] and have applications in cell−substrate studies, cell microenvironment modulation, chemical gradients on surfaces for motility assays, protein−protein interaction studies, creation of diverse libraries for drug screening and toxicology studies, screening of multiple biomarkers in point-of-care personalized medicine, for example. Established biopatterning methods locally deposit analytes using minute volumes (picoliter to microliter) of reagents and can broadly be classified into two categories. The first one uses inkjet technologies, where nanoliter volumes are spotted onto surfaces.[6−8] The second category requires a gentle contact between a pin and a substrate[9] to transfer a small volume of processing liquid onto a surface. Both approaches (Figure 1a) are widespread in research laboratories and industrial facilities as they enable high-throughput processing and precise (nanometer to micrometer accuracy) deposition[10,11] of biochemicals. However, these approaches are limited by uncontrolled wetting and evaporation,[12] which affect the homogeneity and repeatability of deposition.[6,13] More generally, to abate evaporation, oil has been used as an immersion liquid,[14] but in the context of biopatterning, the surface requires a rigorous wash step to remove the oil prior to downstream analytical tests. Such rinsing involves solvents and surfactants that will likely cause degradation of the patterned receptors. In contrast, several research groups developed microfluidic-based biopatterning techniques focusing on deposition quality[15] in which closed channels prevent evaporation. For example, Delamarche et al. developed microfluidic networks (MFN, Figure 1c) to deliver proteins to surfaces by placing and sealing elastomeric materials on the substrate[16] and, a variant thereof, a stencil-based method[17] to spatially localize the processing liquid on surfaces (Figure 1b). These microfluidic methods resulted in high-quality biopatterns confined to specific areas on a surface but suffered from either a large volume consumption or a low deposition rate. Moreover, MFNs are not compatible with high-density discrete unit patterns, such as microarrays, and any variations of the pattern would need a redesign of the network. Other examples of contact-based microfluidic implementations, such as chemistrodes,[18] fountain pens,[19,20] and continuous-flow printing,[21] also impose constraints on the type of surface and the ability to scan and are subject to cross-contamination as well. Noncontact implementations using electric fields, such as electrohydrodynamic jet printing[22] and scanning ion-conductance microscopy,[23] demonstrated patterns in the hundreds of nanometers range with large inter- and intraspot variations[24] but require conductive substrates.

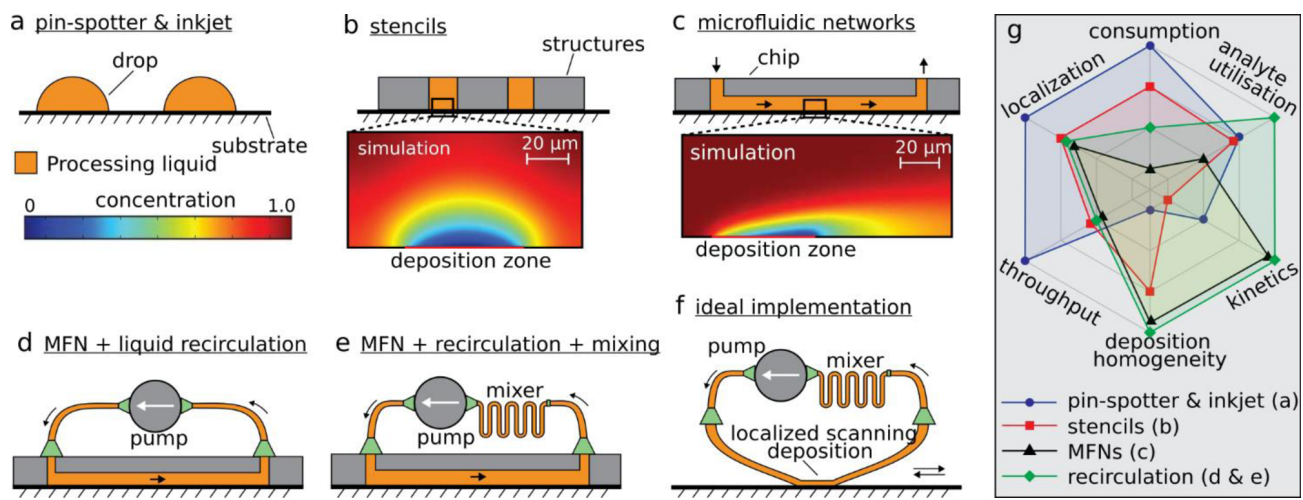

**Figure 1.** Overview of biopatterning methods. (a) Pin-spotter and inkjet deposit small volumes of processing liquid on substrates. (b) Stencils use structures that are physically placed on a substrate to localize the processing liquid with micrometer precision. The simulation shows the diffusion profile above the deposition zone. (c) Microfluidic networks (MFN) use convection-enhanced deposition to decrease deposition time and a microchannel for spatial localization. The simulation illustrates the diffusion profile along the flow path. (d) Using a pump to recirculate the processing liquid enables a more efficient usage of the processing liquid. (e) Mixing of the recirculated liquid is important to increase usage efficiency in microfluidics, where flows typically are laminar. (f) The ideal implementation of biopatterning takes advantage of convection-enhanced deposition, high-resolution scanning, recirculation, and mixing while ensuring a localized, noncontact deposition. (g) Empirical graph highlighting six important attributes of biopatterning methods.

Thus, versatile and high-quality patterning of biochemicals on surfaces remains elusive, but microfluidic implementations have paved the way toward convection-enhanced deposition. In general, continuous-flow methods (Figure 1c–f) result in a reduction of the deposition time compared with diffusion-driven processes[25] but are very inefficient in terms of reagent consumption for two key reasons. First, continuous flow implies the use of larger volumes than in diffusion-based deposition, and second, the actual usage of analytes from the solution is very low. For example, a typical convection-based surface reaction in a 100-$\mu$m-deep channel would consume less than 1.5% of the sample flowing over the surface, resulting in tremendous waste of analytes (see section SI-1 in the Supporting Information). This inefficient reagent usage is problematic, particularly in biopatterning, where biochemicals such as antibodies and DNA probes are expensive. Circulating the processing liquid multiple times over the deposition zone (Figure 1d) provides a way to improve reagent utilization.[26,27] In microchannels, however, laminar flows will hinder homogenization of the recirculated volume, and mixing would therefore be necessary to enable a more efficient usage of the processing liquid (Figure 1e).[28,29] With this in mind, we envision six attributes for an ideal implementation of a versatile biopatterning device (Figure 1g): low reagent consumption, high deposition rate, efficient reagent usage, low variation in the spots deposited, high throughput, and micrometer-scale precision in deposition. Such a device should leverage continuous flow deposition together with mixing and recirculation of the processing liquid (Figure 1f). On the one hand, this ideal implementation would retain the advantages of inkjet and pin-spotting devices, namely, low reagent consumption, high throughput, and precise localization. On the other hand, processing of the surface with continuous flows would provide efficient use of analytes, reduced deposition time, and homogeneity of the pattern deposited.

In this paper, we report a scanning, noncontact microfluidic device for high-quality, versatile biopatterning. This approach makes use of a vertically oriented microfluidic probe (MFP)[30,31] to confine nanoliter volumes of processing liquid on top of a substrate in a wet environment, ensuring noncontact operation and convection-enhanced deposition. MFPs belong to a class of devices termed "open-space microfluidics" that relies on hydrodynamic flow confinement (HFC) of the liquid.[32] Here, we leveraged hierarchical HFC[33] to dilute the aspirated processing liquid minimally and show that biochemicals can be efficiently recirculated back and forth on a surface. We developed analytical models to investigate diffusion, advection, and surface reactions in the context of HFC and, with this, defined the MFP operational parameters for efficient biopatterning (see section SI-2 in the Supporting Information). We demonstrate the deposition of analytes in the context of the deposition and detection of IgG, showing a 2- to 5-fold increase in deposition rate together with a 10-fold reduction in analyte consumption while ensuring less than 6% variation in pattern homogeneity on a standard biological substrate. IgG antibodies are macromolecules of about 150 kDa and play a key role in the immune system as they react with receptors present on the surfaces of macrophages, neutrophils and natural killer cells. They are used as a diagnostic marker for several autoimmune diseases and as a measure of the immune response to pathogens, for example, the serologic immunity to measure measles, mumps, rubella, hepatitis B, varicella. IgGs extracted from donated plasma are also used in therapy to treat immune deficiencies, autoimmune disorders, and infections.[34]

In addition, we demonstrate recirculation of a processing liquid using the MFP in the context of a surface assay: (i) for probing 12 independent areas with a single microliter of processing liquid, which is relevant when multiple assays have to be performed with a limited volume budget, and (ii) for processing a 2 mm² surface to create 170 antibody spots of 50 × 100 $\mu$m² area using 1.6 $\mu$L of liquid.

For efficient usage of the processing liquid, we circulate this liquid back and forth while ensuring homogenization through mixing in the serpentine channels of the MFP head.[29] The implementation of this circulation in the device has two states, in which the direction of flow is different. In state 1, the liquid is injected via the two apertures on the right and aspirated through the two apertures on the left (see Figure 2a). After switching to state 2, all flow directions of the liquids are inverted. Such liquid switching would typically be performed with 2-position valves, but they generally have a large dead volume (tens of microliters) and involve a displacement of the liquid that would disrupt the HFC temporarily. To address these two important issues, we developed a low-dead-volume fluidic system in which the liquid reservoirs are placed close to the MFP head (see Figure 2b). We used the valves to redirect positive and negative pressures toward the appropriate reservoirs. The advantages of using valves to switch the pressure instead of liquids are (i) the absence of liquid displacement in the HFC when switching, (ii) valves can be placed away from the microfluidic system without increasing dead volumes, and (iii) switching is fast (within few milliseconds), with the pressure stabilizing within 1 s. Flow rates are generated using external pressure controllers and hydrodynamic resistors (see Figure 2) integrated in the MFP head and monitored in real-time.

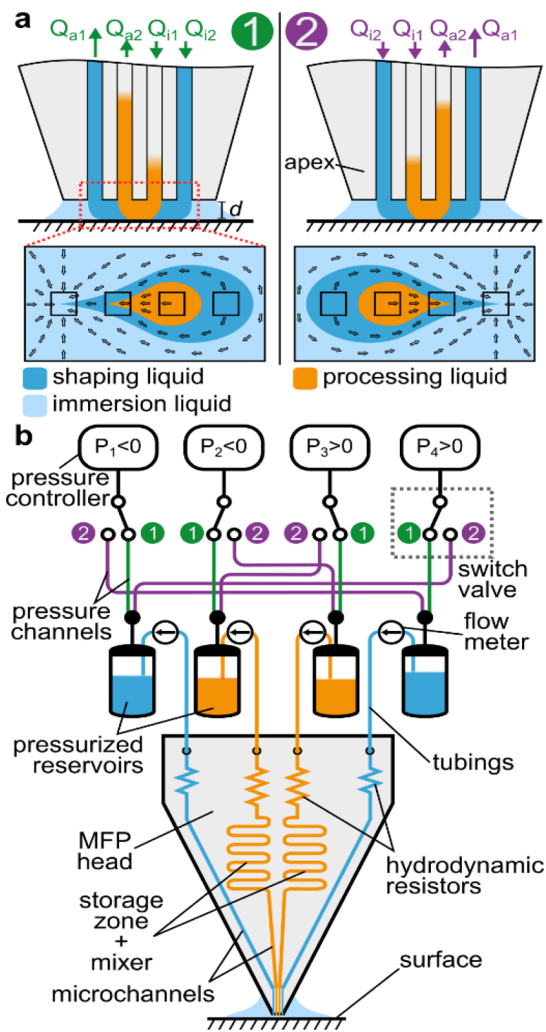

Figure 2. Scheme for liquid recirculation using pressure switching. (a) Hierarchical flow confinement is generated in state 1 to ensure minimal dilution of the processing liquid. When switching the valves to state 2, the direction of the flows is reversed. (b) Scheme of low-dead-volume pressure switching for liquid recirculation.

## ■ THEORY

**Diffusive Transport between Two Laminar Flows within the HFC.** In a single HFC, dilution of the processing liquid is driven primarily by its aspiration together with the surrounding liquid. In the hierarchical HFC, because $Q_{i1} = |Q_{a2}|$, dilution is solely due to diffusion of analytes from the processing liquid to the shaping liquid and is therefore limited. For efficient recirculation of the processing liquid, this loss of analyte should be minimized. We developed a model to investigate the dilution $\gamma$ of the processing liquid as a function of two key parameters, namely, the apex-to-surface distance $d$ and the flow rate of the processing liquid $Q_{i1}$ (see Figure 3a,b).

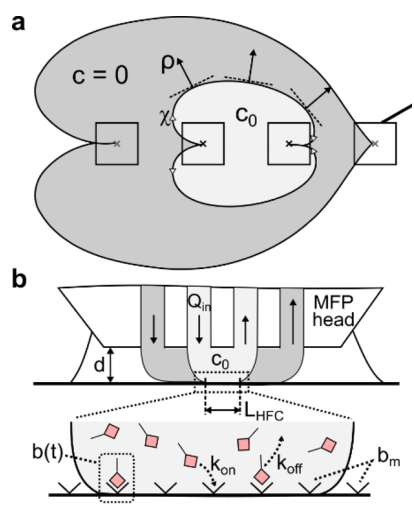

Figure 3. Scheme for analytical model of analyte transport in a hierarchical HFC. (a) Bottom view of a hierarchical HFC with corresponding coordinate system. (b) Schematic side view of a hierarchical HFC.

The apex of the MFP head and the surface to be processed are considered as two parallel surfaces, with an apex-to-surface distance ranging from 10 to 50 μm. Given these boundary conditions, we apply the Hele−Shaw approximation and model the liquid flow between apex and surface as a potential flow (see section SI-3 in the Supporting Information). Each of the four apertures creates a static, radial velocity field of liquid flow. The resulting velocity field represents the superimposition of the radial velocity fields from the individual apertures. Because the resulting velocity field is symmetric, transport across the interface can be described by considering only one-half of the interface. We define a curvilinear coordinate system, with $\chi$ the coordinate axis tangential to the interface and $\rho$ the coordinate axis perpendicular to the interface (see Figure 3a). The analyte in the processing liquid has a diffusion coefficient $D$, and the initial analyte concentrations are $c_0$ and $c = 0$ for the processing liquid and the shaping liquid, respectively. The initial concentration profile $c(\chi = 0, \rho)$ across the interface can thus be approximated with a Heaviside step function.

The dilution $\gamma$ is expressed as the ratio of the rate with which analytes diffuse from the processing liquid into the shaping liquid to the total rate of analytes transported through the confined liquids:

$$\gamma = \frac{\partial n_D}{\partial t} \bigg/ \frac{\partial n_{in}}{\partial t} = \frac{2d}{Q_{i1}} \int_0^{\chi_{max}} \sqrt{\frac{D \cdot \hat{v}_\chi}{4\pi u}} \, du \quad (1)$$

Theoretical values for $\gamma$ and the development of equations are presented and compared with experimental dilution values in section SI-3 in the Supporting Information.

According to both analytical and experimental dilution values, efficient recirculation is favored at higher flow rates and when the head is in close proximity of the surface. We note however that the flow rate will influence the amount of processing liquid used per circulation cycle. This implies that for a finite volume of processing liquid, there is a trade-off between the flow rate of the processing liquid and the number of circulation cycles per minute. Depending on the application, parameters such as apex-to-surface distance and flow rates need to be adjusted to ensure proper surface processing and minimal loss of processing liquid.

**Convective Transport and Deposition of Analytes on a Surface.** This model describes a metric $\varepsilon(t)$ that quantifies the benefit of convective transport as compared to diffusion-driven transport for surface biopatterning. While it is clearly established[25] that convection will enhance the deposition rate, this gain largely depends on the working regime and parameters such as analyte concentration in the processing liquid, flow rates, and surface processing duration. We developed an analytical model that describes convection-enhanced deposition of analytes on a surface using the hierarchical HFC. This model accounts for the transport of IgG molecules from the HFC to the surface and for the kinetics of the reaction between analytes and receptors on the surface.

We investigate the deposition of an analyte with a diffusion coefficient $D_A$ and a concentration $c_0$ on a surface presenting binding sites with a surface density $b_m$. Binding of analytes is assumed to follow first-order Langmuir kinetics and can therefore be characterized with the association and dissociation constants $k_{on}$ and $k_{off}$. Because of the binding of analytes to receptors on the surface, the concentration of analytes in the processing liquid directly above the surface reduces, and a depletion zone is formed. Despite strong advective transport, analytes can travel across the depletion zone only by means of diffusion. For operation of the MFP, this depletion zone is steady and on the order of a tenth of a micrometer, which is small relative to the apex-to-surface distance and to the HFC footprint, which is typically $50 \times 100$ $\mu m^2$. We apply models for transport through a thin depletion zone and first-order binding of analytes to a surface as summarized by Squires et al.[25] Central to this analysis is a Damköhler number $Da$ defined as the ratio of the rate of reaction at the surface to the rate of convective transport of analytes.

$$Da = \frac{c_0 k_{on} b_m}{N \mathcal{F} D} \quad (2)$$

In eq 2, $N$ is Avogadro's number and $\mathcal{F}$ is the non-dimensionalized flux of analytes through the depletion zone.[35] For MFP-based deposition of an IgG, the binding of analytes to the surface is neither transport nor reaction limited, because $Da$ is in the range of 1 (see section SI-4 in the Supporting Information). The ratio $\varepsilon(t)$ of analyte bound with the MFP to that of pipet deposition can be evaluated as (see section SI-4 in the Supporting Information):

$$\varepsilon(t) = \frac{(1 - e^{-k_{on}c_0 + k_{off}/Da \cdot t})}{(1 - e^{-(k_{on}c_0 + k_{off})40 D_{Ig}N/k_{on}b_m L_{HFC}\sqrt{t}})} \quad (3)$$

Here $\varepsilon(t)$ expresses the ratio of MFP-deposited IgG to pipet-deposited IgG as a function of time. This metric quantifies the benefit of convective transport when compared with diffusion-driven surface patterning. Figure 4a shows $\varepsilon(t)$ for four concentrations of IgG in the processing liquid, with standard parameters for MFP surface processing. The graph suggests that deposition using the MFP is more efficient for times shorter than 10 min and strongly depends on the analyte concentration. Once saturation of the surface with the MFP has been reached, the pipet deposition efficiency will slowly converge to the MFP-based deposition efficiency, thus $\varepsilon(t)$ converges to 1. Interestingly, the lower the concentration, the better the MFP will perform in comparison to pipet deposition, which implies that convective deposition is particularly favorable for low concentrations. As an example, for a concentration of 50 $\mu g/mL$, the amount of analyte deposited with the MFP after 40 s will be 1.5-fold higher than with pipet deposition. Longer deposition times will result in identical deposition efficiencies for both approaches after 6 min. In contrast, for a 10-fold lower concentration (5 $\mu g/mL$), the number of analytes deposited with the MFP after 5 min will be 3.5-fold higher than with pipet deposition. The pipet deposition will require more than an hour to reach the MFP-based deposition efficiency. This analysis clearly highlights the Analytical ChemistryArticleadvantage of using MFP deposition for low concentrations of analyte and also that there is an optimal range of processing durations in which the MFP is particularly relevant.

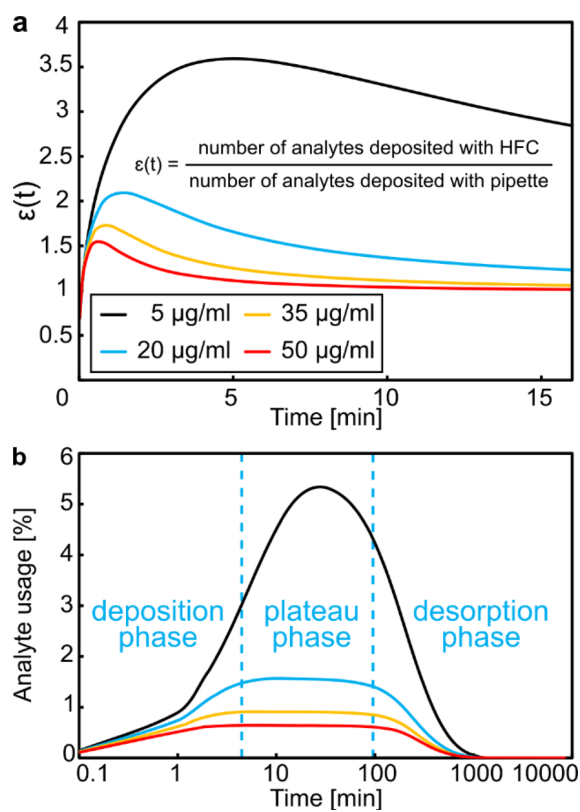

Figure 4. Deposition efficiency using the MFP with different concentration of reagents. (a) Ratio of deposited analytes with the MFP compared with pipet-based deposition for concentrations of 5, 20, 35, and 50 $\mu g/mL$. (b) Ratio of analytes deposited over the total amount of analytes in the processing volume. Both values were calculated numerically for four concentrations of an IgG molecule.

From this model, we also derived the analyte usage as a function of time (see Figure 4b), which corresponds to the ratio of analytes bound to the surface to the initial number of analytes in the processing liquid. Consequently, analyte usage is first marked by a strong increase in the number of bound analytes, and we term this the deposition regime. The decrease of available free binding sites on the surface over time leads to a reduction of the association rate, and analyte usage enters a plateau regime when the association rate and the dissociation rate balance each other. Through successive dilution of the processing liquid in every circulation cycle, the concentration of analytes decreases to the extent that dissociation of analytes from the surface becomes predominant in the desorption regime. The model allows us to estimate the number of recirculation cycles (or time) after which the processing liquid is depleted because of deposition on the surface and diffusive transport to the outer flow confinement. The processing liquid can then be replenished before entering the "desorption phase", thus preventing deposition issues. Figure 4b further implies that analyte usage depends on the initial concentration and remains below 6% for the four concentrations investigated. An important implication of this result is that a unit volume of processing liquid can be circulated multiple times for multispot deposition or for long incubation times.

## ■ MATERIALS AND METHODS

**Microfluidic Probe Platform.** The MFP platform consists of three linear stages (Lang GmbH, Hüttenberg, Germany) for the positioning of the head relative to the surface. The sample is placed on a custom arm of the stages that is placed above an inverted microscope (Eclipse TI-E, Nikon, Japan). The head can be positioned on the surface with an accuracy of 100 nm. Images were acquired using an ORCA-Flash 4.0 camera (Hamamatsu Photonics K.K., Hamamatsu, Japan) and using LED lamp illumination (Sola, Lumencore Inc., Beaverton, OR). The MFP head is mounted on a holder vertical to the microscope objective and aligned to the surface using manual rotation stages. A detailed description of the setup, alignment procedure, and standard operation of the MFP head can be found elsewhere.[33]

**Setup to Implement Liquid Recirculation.** Four independent reservoirs were connected to the MFP head using 1/16 PEEK tubing (IDEX H&S, Oak Harbor, WA) and linear connectors (Dolomite Microfluidics, Charlestown, MA). The reservoirs were pressurized using pressure control devices (MFCS, Fluigent, Paris, France) with a working range of ±200 mbar per channel. Pressure was switched in the four reservoirs simultaneously using two-way switch valves (2-switch, Fluigent, Paris, France). Pressures would typically stabilize within hundreds of milliseconds, resulting in the stabilization of the flow confinement within one second. Flow rates were measured using flow sensors (Fluigent, Paris, France). Hydrodynamic resistances in the MFP head were designed to provide a suitable working range of pressures to generate the flows.

**MFP Head Microfabrication.** The MFP head is a microfabricated silicon-glass device. The head comprises four apertures, hydrodynamic resistors, and two storage and mixing zones of 1 μL volume for circulation of the processing liquid. Channels (50 μm × 50 μm) in silicon were defined photolithographically and etched using deep reactive-ion etching (DRIE). Subsequently, they were sealed by anodic bonding (1.3 kV, 475 °C) with glass. The fabrication process has been described in detail elsewhere.[31]

**Measuring the concentration of processing liquid.** We used a solution of 50 μM Rhodamine B as the reference for concentration measurements. The fluorescence intensity of the flow confinement was acquired using a camera. We measured the average fluorescence value on a region of interest (ROI). To obtain the relative drop in concentration after each circulation cycle, the same ROI was used to measure the fluorescence of the confined solution. For measurements requiring precise apex-to-surface distance control, we performed a $z$-axis reference to zero height prior to each experiment.

**Rabbit IgG Antibody−Antigen Assay.** A test surface was prepared starting with a clean polystyrene Petri dish that was incubated 30 min at room temperature with 100 μL of a 50 μg/mL rabbit IgG solution (Sigma-Aldrich, Saint Louis, MO). After three rinsing steps (PBS + Tween-20 0.05, PBS, and DI water), the surface was blocked for 30 min with BSA (1% in PBS) and rinsed. Fluorescently labeled rabbit anti-IgG was deposited on the surface with the MFP for different deposition times.

**Automated Deposition of Antigens on a Surface with the MFP.** After priming the fluidic tubing with the appropriate solutions (fluorescent anti-IgG in the inner injection, PBS everywhere else), a Matlab script was used for automating the sequential surface processing steps: stage movement for precise spotting, and valve switching for recirculation and pressure control. Pressure controllers and switch valves were controlled by the manufacturer's API and the stages with a standard serial protocol.

**FEM Modeling.** Steady-state 2D transport simulations were performed using COMSOL Multiphysics version 4.2. We used incompressible fluids, open boundaries, and nonslip conditions on surfaces as parameters. The model couples the solution of the Navier−Stokes equation and the convection-diffusion equations. All liquids were chosen to be water (incompressible Newtonian fluid with a density of 998 kg/m$^3$ and a dynamic viscosity of 0.001 N s/m$^2$). The diffusion constant of the analyte was set to $D = 3.8 \times 10^{-7}$ cm$^2$/s, which corresponds to the diffusion constant of an IgG molecule.

**Image Analysis.** Time-lapse images were analyzed using ImageJ and NIS Elements Basic Research software suite (Nikon). Scale bars were calculated from bright-field images. To account for a potentially uneven illumination pattern, each image was divided by a control image taken on an empty zone of the slide. After removal of the background, the processed image was used to calculate the fluorescence intensity.

**Statistical Analysis.** Error bars represent the standard deviation. If no error bars are visible, the standard deviation is smaller than the symbol representing the mean value. "$n$" refers to the number of data points unless specified otherwise.

## ■ RESULTS AND DISCUSSION

To implement an MFP-based antibody/antigen assay, we used a standard polystyrene Petri dish as substrate. After incubation of IgG, we processed the surface with the MFP using a solution of fluorescently labeled anti-IgG for different durations with a constant flow rate (1 μL/min) and an apex-to-surface distance $d = 30$ μm. The results of these assays ($n = 5$ experiments per deposition time) are presented in Figure 5. The experiments were performed both with and without recirculation of the

3239

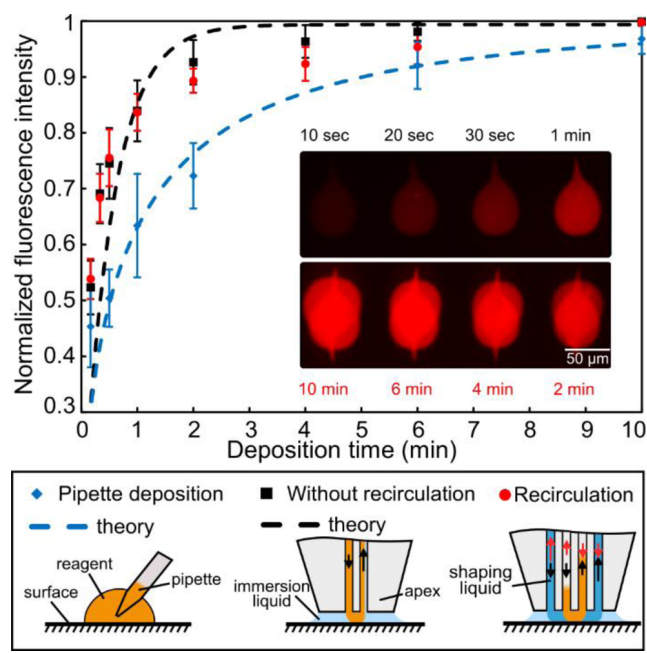

**Figure 5.** Deposition of goat anti-IgG on IgG-coated polystyrene surface. Normalized fluorescence intensity of deposited anti-IgG ($n = 5$) and corresponding trends from the analytical model (theory). Error bars represent the standard deviation for five experiments. The fluorescence intensity is measured in the center (5 $\mu$m × 5 $\mu$m) of the teardrop-shaped HFC footprint.

processing liquid, along with a reference experiment using pipet deposition. The fluorescence intensity of deposited anti-IgG was used to assess the deposition efficiency and quality. Deposition using the MFP (Figure 5, black squares and red dots) showed a higher efficiency than pipet deposition (Figure 5, blue diamonds). With an antigen concentration of 50 $\mu$g/mL, we saturated the surface in less than 3 min with the MFP and in 6 min with the pipet. Interestingly, we observed no significant difference between deposition with and without recirculation, as both methods use convection-enhanced deposition. However, the model for convective deposition predicts saturation to occur faster than what we observed in the experiments. We hypothesize that these discrepancies arise from (i) three-dimensional effects that are not accounted for in the model, resulting in a reduction of the effective flow velocity and therefore a reduction of advective transport to the surface, and (ii) the fact that $k_{on}$, $k_{off}$, and the binding sites surface density $b_m$ were derived from the diffusion-driven deposition experiments (see Figure 5, blue line and symbols) and may be different for the MFP-based deposition experiments. A striking example of the benefits of recirculation using the MFP is that the total volume used after 10 min with recirculation (1 $\mu$L) was 1 order of magnitude lower than without recirculation (10 $\mu$L), while reaching an identical density of captured antigens. Ultimately, this approach would lead to either a drastic reduction of antigens needed for the assay or, conversely, a reduction of the assay time if a preconcentration of the analyte is done prior to recirculation. Moreover, when using recirculation, the total volume of analyte needed is largely independent of the reaction duration. This result is particularly relevant in the case of low-concentration analytes, where the reaction time can be in the range of hours. Recirculation therefore enables enhanced kinetics as the result of convection, while improving reagent usage through multiple circulations of the same volume.

Most biopatterning applications require deposition on multiple zones with high spatial resolution, typically in the micrometer range. A key advantage of the MFP is its capacity to interact locally with a surface and to scan large areas rapidly.[36] Leveraging this capacity, we investigated the recirculation of a given volume of liquid on multiple positions on a substrate. Multizones deposition can be used for localized capture of antigens from a sample, and using a single microliter, recirculation allows multiple independent capture zones to be probed. Similarly, an unknown concentration of antigen can be recirculated for different lengths of times on multiple areas on an antibody-coated surface. This can be leveraged to determine the adequate deposition time on the surface to prevent over- and under-exposure of the sample of interest in the capture zone. We demonstrated these two aspects by recirculating different concentrations of a fluorescently labeled IgG on an anti-IgG-coated surface (see Figure 6a.) and checked the deposition density after different deposition times (see Figure 6b).

These results indicate that an adequate deposition time (signal intensity detectable and no saturation) for the highest concentration was obtained in under 1 min, whereas the lowest one would require 10 min to yield a signal sufficient for analysis. This use of recirculation is a unique implementation of a "reverse assay" in which the sample is processed successively on multiple zones. These independent zones may contain different capture antibodies, giving access to multiple assays with a minimal volume or, for a single antibody, enable better control of the capture density and, ultimately, quantification of antigens concentration through multiple deposition durations.

We also investigated the possibility of using a limited and small volume of processing liquid in the MFP head to deposit a large number of spots on the surface. We deposited 170 spots (see Figure 6c), switching the flow direction every ten spots, with the probe staying 10 s on top of each spot. Working close to the surface (10 $\mu$m), we limited the dilution to the extent that only a minimal decrease in fluorescence was observed between the first and the last spot (Figure 6d). This decrease in fluorescence was predicted by combining the two analytical models describing dilution during recirculation and binding to the surface. The models exhibited very good correlation with the experimental values. A volume of 1.66 $\mu$L of fluorescently labeled IgG were used to perform deposition of these 170 spots, so each spot corresponds to a "used" volume of 9.7 nL.

Finally, we investigated the quality of the spots deposited and found intra- and interfootprint variations of 5.8% and 3.4%, respectively (see Figure 5e,f), confirming the advantages of "wet" methods for presenting an antibody to a surface in terms of deposition quality. We hypothesize these relatively small deposition variations to be due to the type of substrate used (standard, untreated polystyrene Petri dish) and that the deposition quality could be further increased by using engineered substrates.

## ■ CONCLUSION

We described, characterized, and validated a new method for surface processing and biopatterning. This method exploits the hierarchical implementation of the HFC to efficiently present micro- to nanoliter volumes of analytes on surfaces. The analytical models we present provide guidelines on minimizing the dilution of the processing liquid and enhancing the deposition kinetics by controlling parameters such as the apex-to-surface distance, the flow rates and the deposition time. This enables the rapid processing of areas ranging from 5 × 5 μm$^2$ to 1 × 1 cm$^2$ in large substrates, such as Petri dishes and glass slides, with volumes in the microliter range. As an example, we deposited 170 antibody−antigen spots with a per-spot volume of 9.7 nL and with a variation in deposition homogeneity below 6%. In comparison, state-of-the-art inkjet tools would require a comparable volume per spot but achieve limited deposition homogeneity. Importantly, the high quality of the patterns deposited with our method is compatible with the needs of quantitative surface assays. Experiments comparing the deposition efficiency showed a 3-fold improvement of the deposition speed and a 10-fold decrease of the total volume needed for a surface assay when using convection-enhanced deposition and liquid recirculation in the MFP. Such outcomes are particularly relevant in the case of expensive biochemicals or when the concentration of analytes is so low that it would impede deposition efficiency. In addition, we demonstrated that this tool can probe multiple capture areas independently with a single limited volume of a biological (clinical) sample.

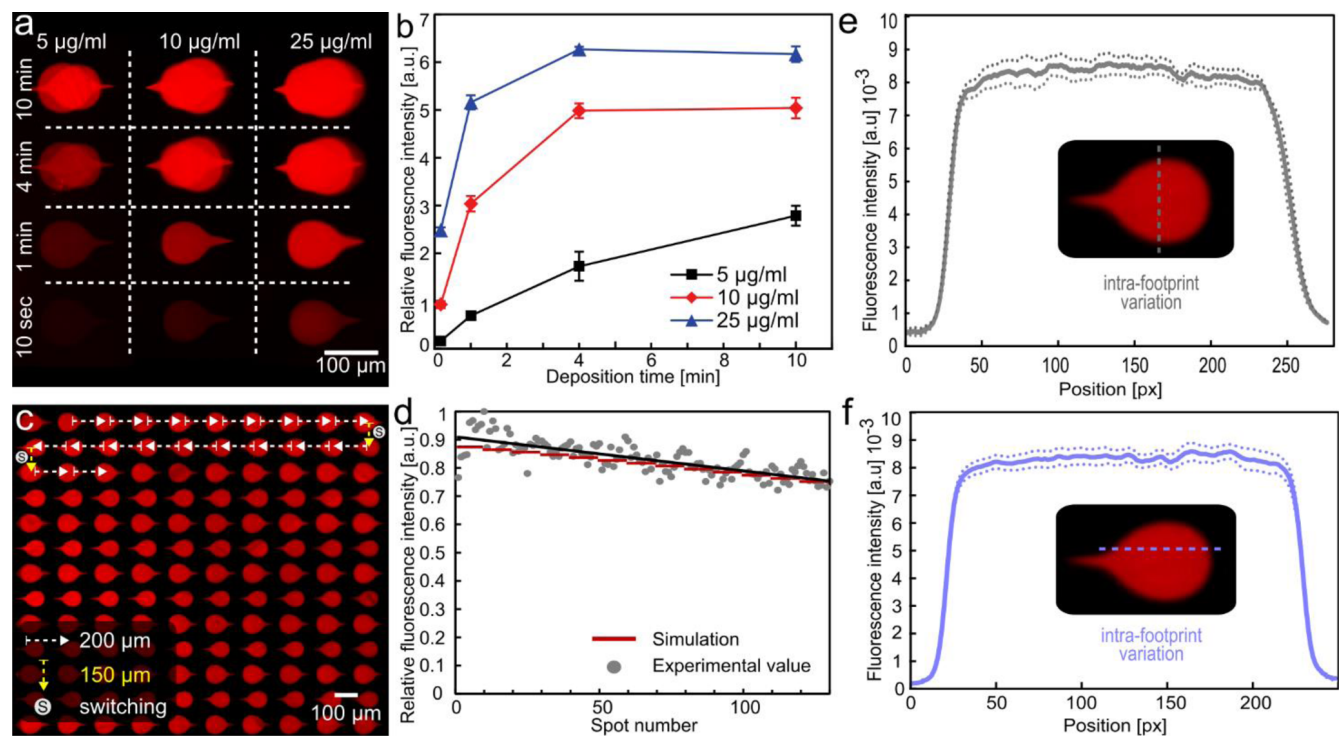

**Figure 6.** Multispot deposition of goat anti-IgG using recirculation with the MFP. (a) Image of the deposition of anti-IgG solution using 1.6 μL with concentrations of 5, 10, and 25 μg/mL and incubation times of 10 s, 1 min, 4 min, and 10 min. (b) Fluorescence intensity of zones shown in part a. (c) Photograph of 130 spots deposited with 1 μL of solution (170 spots in total). The flow direction was switched every ten spots. (d) Average fluorescence intensity measured on each spot compared with the predictions of our models. (e and f) Intrafootprint variation within one spot (<6% variation).

For efficient biopatterning, we leveraged the strengths of the MFP and the underlying physics of HFC. First, the hydrodynamic confinement of the processing liquid with the MFP eliminates the need for microfluidic networks and channels or multiphase systems for localization and compartmentalization. The MFP provides a dynamic and highly versatile means of localizing and presenting the processing liquid on a surface immersed in an aqueous environment. This method is compatible with Petri dishes, glass slides, as well as more challenging substrates with varying topologies, such as cell cultures and tissues sections. Second, this method can be scaled to form larger or smaller footprints by changing the aperture size, aperture spacing, and flow rates with dimensions ranging from hundreds of nanometers to a few millimeters. The scanning capability of the probe[36] allows the patterning on large surfaces, typically in the centimeter range, with the possibility of creating arrays of up to 30 000 spots/cm$^2$ with a spot size of 10 × 10 μm$^2$. Third, the capacity of the MFP to rapidly switch liquids[37] on top of the substrate makes it a relevant tool for implementing complex biochemistries requiring multiple consecutive chemicals to be dispensed on the surface or even for manufacturing protein microarrays.

A current limitation of the device is its throughput. In this paper, a single spot is patterned at a given time, but we foresee processing multiple positions on the surface in parallel by leveraging standard microfluidic channel bifurcations implemented on the probe. We believe that the MFP might not be the preferred tool when the incubation or reaction time is in the range of tens of hours. However, as discussed in this paper, most surface chemistries will be accelerated by convective transport, and thus long overnight incubation will likely not be required. For specific applications, increasing the recirculated volume is readily feasible, but working with processing liquid volumes below 100 nL is challenging to implement in the current probe configuration. However, we believe that the microliter to milliliter range is appropriate for most biopatterning applications, as it is compatible with standard volumes used in industrial methods, such as inkjets and pin-spotters. Finally, the asymmetric teardrop shape of the deposition area can be altered by using different aperture spacings and geometries. The method presented in this paper combines five of the six attributes of an ideal biopatterning approach, namely, low reagent consumption, high deposition rate (kinetics), efficient reagent usage, low variation in the spots deposited, and micrometer-scale precision in deposition. We strongly believe that this combination of advantages creates a powerful tool for an efficient and high-quality patterning of receptors on surfaces and thus will enable quantitative assays in discovery research, point-of-care devices, large-scale surface patterning, and reverse immunoassays and will catalyze the manufacturing of protein microarrays. This method will be a unique facilitator in quantitative biology and precision diagnostics.

## ■ ASSOCIATED CONTENT

### ● Supporting Information

The Supporting Information is available free of charge on the ACS Publications website at DOI: 10.1021/acs.analchem.5b04649.

> Analyte consumption in a microchannel, operation of the microfabricated MFP head for liquid recirculation, diffusive transport between two laminar flows within the HFC, deposition of analytes on a surface using the MFP (PDF)
>
> Matlab code for estimation of flow velocity fields (TXT)

## ■ AUTHOR INFORMATION

### Corresponding Author
*E-mail: gov@zurich.ibm.com.

### Author Contributions
†J.A. and J.F.C. contributed equally yo this work.

The manuscript was written through contributions of all authors. All authors have given approval to the final version of the manuscript.

### Notes
The authors declare no competing financial interest.

## ■ ACKNOWLEDGMENTS

This work was supported by the European Research Council (ERC) Starting Grant, under the 7th Framework Program (Project No. 311122, BioProbe). We thank Emmanuel Delamarche, Aditya Kashyap, and Prof. Moran Bercovici (Technion) for valuable scientific discussions, Ute Drechsler and Yuksel Temiz for their help in MFP head fabrication, Robert Lovchik for technical assistance, and Prof. Bradley Nelson (ETH Zurich), Prof. Philippe Renaud (EPFL), Bruno Michel, and Walter Riess are acknowledged for their continuous support.